\documentclass[prl,superscriptaddress,amsmath,amssymb,aps,twocolumn,showpacs]{revtex4}
\usepackage[]{inputenc}
\usepackage[dvips]{graphicx}
\usepackage{color}
\usepackage{bm}
\usepackage{mathpazo}
\usepackage{slashed}
\usepackage[ps2pdf,bookmarks=true,colorlinks,linkcolor=blue,urlcolor=green,citecolor=red]{hyperref}

\renewcommand\a{\alpha}
\renewcommand\b{\beta}
\renewcommand\d{\delta}

\renewcommand\l{\lambda}
\renewcommand\r{\rho}

\renewcommand\t{\tau}

\renewcommand\c{\chi}
\renewcommand\j{\psi}

\newcommand\e{\epsilon}
\newcommand\g{\gamma}

\newcommand\m{\mu}
\newcommand\n{\nu}

\newcommand\p{\pi}

\newcommand\s{\sigma}
\newcommand\f{\phi}

\renewcommand\L{\Lambda}

\renewcommand\O{\Omega}

\newcommand\G{\Gamma}

\newcommand{\fig}[1]{Fig.~\ref{#1}}
\newcommand{\eq}[1]{Eq.~(\ref{#1})}

\newcommand\lb{\left(}
\newcommand\rb{\right)}
\newcommand\ls{\left[}
\newcommand\rs{\right]}

\newcommand{\lan}{\langle}
\newcommand{\ran}{\rangle}

\newcommand\ra{\rightarrow}

\newcommand{\non}{\nonumber\\}
\newcommand\pt{\partial}

\newcommand{\cl}{{\cal L}}

\newcommand{\diag}{{\rm{diag}}}

\newcommand{\Tr}{{\rm Tr}}

\newcommand{\im}{{\rm{Im}}}

\newcommand{\bB}{{\vec B}}
\newcommand{\bE}{{\vec E}}

\newcommand{\jb}{{\bar \j}}

\renewcommand{\part}{{\rm part}}
\newcommand{\tr}{{\rm tr}}

\renewcommand{\vec}{\boldsymbol}
\newcommand{\be}{\begin{equation}}
\newcommand{\ee}{\end{equation}}
\newcommand{\bear}{\begin{eqnarray}}
\newcommand{\eear}{\end{eqnarray}}
\newcommand{\ba}{\begin{array}}
\newcommand{\ea}{\end{array}}

\begin{document}

\title{Electromagnetic triangle anomaly and neutral pion condensation in QCD vacuum}
\author{Gaoqing Cao}
\affiliation{Physics Department and Center for Particle Physics and Field Theory, Fudan University, Shanghai 200433, China.}
\author{Xu-Guang Huang}
\affiliation{Physics Department and Center for Particle Physics and Field Theory, Fudan University, Shanghai 200433, China.}

\date{\today}

\begin{abstract}
We study the QCD vacuum structure under the influence of an electromagnetic field with a nonzero second Lorentz invariant $I_2=\bE\cdot\bB$. We show that the presence of $I_2$ can induce neutral pion ($\p^0$) condensation in the QCD vacuum through the electromagnetic triangle anomaly. Within the frameworks of chiral perturbation theory at leading small-momenta expansion as well as the Nambu--Jona-Lasinio model at leading $1/N_c$ expansion, we quantify the dependence of the $\p^0$ condensate on $I_2$. The stability of the $\pi^0$-condensed vacuum against the Schwinger charged pair production due to electric field is also discussed.
\end{abstract}

\maketitle

\section{Introduction}
The vacuum structure of Quantum Chromodynamics (QCD) has been intensively studied over many years and a variety of interesting properties have been extracted. For example, it is well-known that the normal QCD vacuum possesses the spontaneous (approximate) chiral symmetry breaking (CSB) which can be characterized by the scalar quark-antiquark condensate $\lan\jb\j\ran\neq0$; while the pseudoscalar condensates like $\lan\jb i\g_5 \t_3\j\ran$ with $\t_3$ the third Pauli matrix are not permitted --- an assertion known as the Vafa-Witten theorem~\cite{Vafa:1984xg}.

Under proper conditions, the QCD vacuum structure can be altered, usually through phase transitions. Examples include the pion superfluid phase at high isospin chemical potential~\cite{Son:2000xc,Kogut:2002zg,He:2005nk} and the color superconducting phase at high baryon chemical potential~\cite{Alford:1997zt,Rapp:1997zu,Alford:2007xm}; in the first case the isospin symmetry is broken while the chiral symmetry is restored and in the second case the color symmetry is broken while the chiral symmetry may or may not be broken.

Strong electromagnetic (EM) fields provide another way to modify the QCD vacuum. Strong magnetic fields may exist in compact stars~\cite{Duncan:1992hi,Thompson:1993hn,Olausen:2013bpa}, heavy-ion collisions~\cite{Skokov:2009qp,Deng:2012pc,Huang:2015oca}, and early universe~\cite{Vachaspati:1991nm,Baym:1995fk,Grasso:2000wj}. Strong electric fields may also be accessible in heavy-ion collisions~\cite{Hirono:2012rt,Deng:2014uja,Voronyuk:2014rna}.
The presence of the magnetic field is known to enhance the scalar $\lan\jb\j\ran$ condensate at zero quark chemical potentials which is called the magnetic catalysis of the CSB~\cite{Gusynin:1994re,Gusynin:1995nb,Miransky:2015ava,Klimenko:1991he}, while the effect of the magnetic field on the CSB at finite temperature or chemical potentials shows novel features which are still not fully understood~\cite{Preis:2010cq,Bali:2011qj,Bali:2012zg,Fukushima:2012kc,Kojo:2012js,Chao:2013qpa}. It was also proposed that when $|e\bB|\gtrsim m_\r^2$ with $m_\r$ the mass of the rho meson, the QCD vacuum will become a superconductor due to charged rho condensation~\cite{Chernodub:2011mc}. The effect of electric field on the CSB was also studied~\cite{Klevansky:1989vi,Babansky:1997zh,Cohen:2007bt} and it was found that the electric field always tends to break the scalar quark-antiquark pairs and thus weaken the CSB.

\begin{figure}[!htb]
\begin{center}
\includegraphics[width=4.5cm]{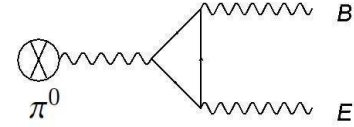}
\caption{The electromagnetic triangle anomaly.}
\label{triangle}
\end{center}
\end{figure}
\begin{figure}[!htb]
\begin{center}
\includegraphics[height=4cm]{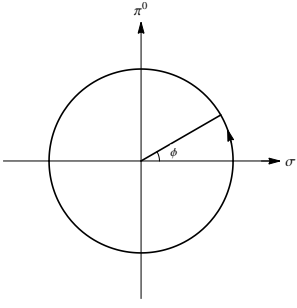}
\caption{Illustration of the chiral rotation due to the neutral pion condensation induced by EM triangle anomaly.}
\label{chiralrotation}
\end{center}
\end{figure}

In this Letter, we study the combined effect of the electric and magnetic fields on QCD vacuum. We focus on the EM-field configuration where the second Lorentz invariant $I_2=\bE\cdot\bB$ is nonzero. We show that the presence of $I_2$ provides a parity-odd environment in which the otherwise-forbidden (by Vafa-Witten theorem) neutral pion condensation can occur via the EM triangle anomaly (see \fig{triangle}). We calculate the $\p^0$ condensate within both the chiral perturbation theory (ChPT) and the Nambu--Jona-Lasinio (NJL) model frameworks and identical result is obtained,
\begin{eqnarray}
\label{pi0cond}
\frac{\p^0}{m^*}=\left\{\begin{array}{ll}{\displaystyle\frac{N_c}{4\p^2 f^2_\p m_\p^2}(q_u^2-q_d^2)\bE\cdot\bB} \;\;\; {\rm for} \,\; |I_2|<I_2^c, \\
{\rm sgn}(I_2) \;\;\;\;\;\;\;\;\;\;\;\;\;\;\;\;\;\;\;\;\;\;\;\;\;\;\;\;\;\;{\rm for}\; |I_2|>I_2^c,
\end{array}\right.\
\end{eqnarray}
where $m^*=\sqrt{(\p^0)^2+\s^2}$ with $\s\sim\lan\jb\j\ran$ and $\p^0\sim\lan\jb i\g_5\t_3\j\ran$ the scalar and neutral pion condensates in the presence of the EM field, $q_u=2e/3$ and $q_d=-e/3$ are charges of $u$ and $d$ quarks with $e>0$ the proton charge, and $I_2^c=4\p^2 f^2_\p m_\p^2/[N_c(q_u^2-q_d^2)]$ is a critical value for $|I_2|$ above which the condensate $m^*$ is wholly contributed by $\p^0$ mode. Equation (\ref{pi0cond}) shows that the $I_2$ rotates the chiral condensate from the $\s$-direction to the $\p^0$-direction with the rotation angle given by $\f=\sin^{-1}(\p^0/m^*)$ if $|I_2|<I_2^c$ and $\f={\rm sgn}(I_2)\p/2$ if $|I_2|>I_2^c$ (see \fig{chiralrotation}).

Throughout this article, we restrict to the zero quark chemical potential and zero temperature case. Before we proceed, we note that the neutral pion condensation we will study is different from the disoriented chiral condensate (DCC)~\cite{Mohanty:2005mv} proposed to possibly occur in heavy-ion collisions, as the latter is a far-from-equilibrium, transient, phenomenon while the former is a nearly equilibrium, (quasi-)stationary, state. We point out that, in the context of DCC, it was also found that the large EM fields in heavy-ion collisions could also possibly catalyze the formation of the DCC~\cite{Minakata:1995gq,Asakawa:1998st}.

\section{Invalidation of the Vafa-Witten Theorem}
The EM-triangle-anomaly induced neutral pion condensation does not contradict the Vafa-Witten theorem. The validity of the Vafa-Witten theorem relies on the positivity, $\det {\cal D}>0$, where ${\cal D}$ is the Euclidean Dirac operator~\cite{Vafa:1984xg,Vafa:1983tf}. However, this is not the case in the presence of a real electric field. To see this, we write down $\cal D$ explicitely,
\begin{eqnarray}
{\cal D}=\g_\m(\pt_\m-ig{\cal A}_\m)+Q\g_4 A_0+iQ\g_i A_i+M,
\end{eqnarray}
where ${\cal A}_\m$ ($\m=1-4$) is the gluon field, $A_\a$ ($\a=0-3$) is the EM field, and $M$ is the mass matrix. We choose the gauge such that $\bE=-{\bm\nabla} A_0$ and $A_i$ is time independent.
For two-flavor case, the charge matrix is $Q/e=\diag(2/3,-1/3)=1/6+\t_3/2$ and the mass matrix is $M=m_0$. The crucial observation is that $Q$ is {\it not} traceless and thus its role is similar with the baryon chemical potential $\m$. It is well-known that the presence of $\m$ destroys the positivity of $\det {\cal D}$, and so does $A_0$. In fact, when $A_0\neq 0$ there does not exist a matrix $\G$ such that $\G^{-1}{\cal D}\G={\cal D}^\dag$ to guarantee the positivity. Choosing another gauge, e.g. $A_0=0$, does not change the conclusion. Because in this case, $A_i$ must depend on time, e.g. for constant $\bE$, $A_i=-E_i t$, which after Wick rotation spoils again the positivity~\cite{footnote}. Note that if $q_u=-q_d$, $\det {\cal D}$ is semi-positive~\cite{Yamamoto:2012bd}.

The invalidation of the Vafa-Witten theorem in the electric field has important physical contents. Physically, in the presence of $\bE$, the Dirac determinant has to be complex or at most semi-positive because it always receives contributions from the Schwinger poles which represent the particle-antiparticle pair creation due to $\bE$~\cite{Schwinger:1951nm}.

\section{Chiral perturbation theory calculation}
We start with the two-flavor ChPT described by the Lagrangian
\begin{eqnarray}
\label{chirall}
\cl=\cl_0+\cl_{\rm WZW},
\end{eqnarray}
where $\cl_0$ is the usual chiral Lagrangian given by (we keep only $O(p^2)$ terms)
\begin{eqnarray}
\cl_0=\frac{f_\p^2}{4}\tr\ls D_\m U^\dag D^\m U+m_\p^2 (U+U^\dag)\rs,
\end{eqnarray}
and the Wess-Zumino-Witten term $\cl_{\rm WZW}$ is given by~\cite{Wess:1971yu,Witten:1983tw,Scherer:2012xha,Son:2007ny,Fukushima:2012fg}
\begin{eqnarray}
&&\cl_{\rm WZW}=\frac{N_c}{48\p^2}A_\m \e^{\m\n\a\b}[\tr\lb QL_\n L_\a L_\b+Q R_\n R_\a R_\b\rb\non&&\;\;\;\;\;\;\;\;\;\;\;\;\;\;\;\;\;\;\;\;\;\;\;\;\;\;\;\;\;\;\;\;\;\;\;\;\;\;\;-iF_{\a\b} T_\n],\\
&&T_\n=\tr\ls Q^2(L_\n+R_\n)+\frac{1}{2}\lb QUQU^\dag L_\n+QU^\dag QUR_\n\rb\rs,\non
&&L_\m= U\pt_\m U^\dag,\non
&&R_\m= \pt_\m U^\dag U.
\end{eqnarray}
In the above, the covariant derivative is given by
\begin{eqnarray}
D_\m U=\pt_\m U+A_\m[Q,U],
\end{eqnarray}
and $U$ is a $2\times2$ unitary matrix representing the chiral fields, for which we choose the Weinberg parametrization,
\begin{eqnarray}
U=\frac{1}{f_\p}(s+i{\bm\t}\cdot{\vec t}),
\end{eqnarray}
where the fields $s$ and ${\vec t}$ fulfill the constraint $s^2+{\vec t}^2=f_\p^2$.

Now we consider constant EM field and make the approximation such that we treat all the chiral fields in (\ref{chirall}) as uniform condensates. We note that the $\s$ and $\p^0$ in \eq{pi0cond} are proportional to $s$ and $t_3$, respectively.
In the absence of the EM field, it is well-known that the CSB takes place along the $\sigma$-direction so that $s=f_\p$ and ${\vec t}={\vec0}$. The EM field with a configuration where $I_2=\bE\cdot\bB\neq 0$ is coupled to the $\p^0$ field via the triangle anomaly and raises the possibility of $\p^0$ condensation. If this happens, the CSB will be rotated from the $\s$-direction to $\p^0$-direction. To explore this possibility, we make the following ansatz for the condensates,
\begin{eqnarray}
s&=& f_\p \cos\f,\;\; t_3=f_\p\sin\f,\\
t_1&=& t_2=0,
\end{eqnarray}
where without loss of generality, we choose $\f\in[-\p/2,\p/2]$ so that $s\geq0$. The Lagrangian $\cl$ then becomes
\begin{eqnarray}
\cl(\f)&=&f^2_\p m_\p^2 \cos\f+\frac{N_c I_2}{4\p^2}\tr(Q^2\t_3) \f,
\end{eqnarray}
where we have omitted a total derivative term. As obviously, the second term is responsible for the triangle anomaly effects such as $\p^0\ra 2\g$. It is easy to minimize $-\cl(\f)$ and obtain
\begin{eqnarray}
\sin\f&=& \frac{N_c I_2}{4\p^2 f_\p^2 m_\p^2}\tr(Q^2\t_3)\non &=&\frac{N_c I_2}{4\p^2 f_\p^2 m_\p^2}(q_u^2-q_d^2), \;\;\;\; {\rm for}\;\; |I_2|<I_2^c,\\
\f&=&{\rm sgn}\lb I_2\rb\frac{\p}{2}, \;\; {\rm for}\;\; |I_2|>I_2^c,
\end{eqnarray}
where $I_2^c=4\p^2 f_\p^2 m_\p^2/[N_c(q_u^2-q_d^2)]$. This is equivalent to \eq{pi0cond}.

Several comments are in order. (1) In the chiral limit $m_\p\ra0$, thus $I_2^c\ra0$, an infinitesimal $I_2$ will drive the maximum $\p^0$ condensation, namely, the CSB is completely driven to $\p^0$-direction once $I_2$ is applied. (2) The neutral pion condensation induced by $I_2$ is merely a chiral rotation; it is not a phase transition as no symmetry is broken along this process. (3) As a low-energy effective theory of QCD, the ChPT is reliable only when all the physical parameters, $m_\p, p, \sqrt{|e\bB|}, \sqrt{|e\bE|}$, are much smaller than the typical hadronic scale $\L_\c\sim 1$ GeV. We can estimate that
\begin{eqnarray}
\frac{e^2I_2^c}{\L_\c^4}\sim \frac{m_\p^2}{\L_\c^2}\ll1,
\end{eqnarray}
which justifies the validity of our result for $I_2$ not too larger than $I_2^c$. (4) As well-known, when $|e\bE|\gg m_\p^2$ (but not larger than $\L_\c^2$), the vacuum will be unstable because of the Schwinger pair production of $\p^+\p^-$. We will discuss this issue in another section.

\section{Nambu--Jona-Lasinio Model calculation}
The ChPT is built on hadronic degrees of freedom, we now use the NJL model which is based on quark degrees of freedom to test the consequences of the ChPT. The Lagrangian of the two-flavor NJL model is given by~\cite{Klevansky:1992qe}
\begin{eqnarray}
\cl_{\rm NJL}=\jb(i\slashed{D}-m_0)\j+G[(\jb\j)^2+(\jb i\g_5{\bm\t}\j)^2],
\end{eqnarray}
where $\j=(u,d)^T$ represents the two-flavor quark fields, $m_0$ is the current mass of quarks, $G$ is the four-fermion coupling constant, and
\begin{eqnarray}
D_\m=\pt_\m+iQA_\m,
\end{eqnarray}
is the covariant derivative. Note that, in addition to the coupling constant $G$, the NJL model contains implicitly another parameter, the ultraviolet cutoff $\L$, which specifies the applicable region of the model. The values of $G$ and $\L$ are determined by fitting NJL predictions with the physical hadronic observables like $f_\p$ and $\lan\jb\j\ran$ in the normal vacuum~\cite{Klevansky:1992qe}.

\begin{figure}[!htb]
\begin{center}
\includegraphics[height=4.5cm]{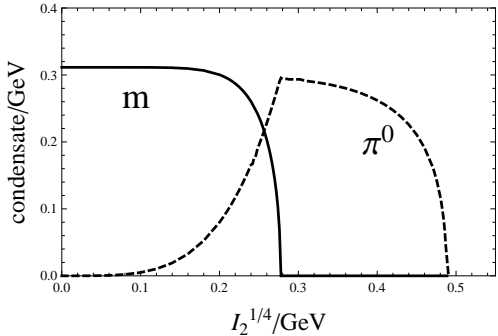}
\caption{The constitute quark mass $m$ and $\p^0$ condensate as functions of $I_2^{1/4}$ obtained by minimizing the thermodynamic potential $\O$ at $I_1=0$ in NJL model. The presence of $I_2$ always tends to diminish $\s$ condensate while drive $\p^0$ condensation.}
\label{fig_m_pi0}
\end{center}
\end{figure}

The NJL model was widely used to study a number of non-perturbative properties of QCD, especially those related to the chiral symmetry. In fact, the NJL model shares the same global symmetries with QCD. At the mean-field level where only the leading-order terms in the $1/N_c$ expansion are kept,
\begin{eqnarray}
\cl_{\rm NJL}=\jb(i\slashed{D}-m-i\p^0\g_5\t_3)\j-\frac{\s^2+(\p^0)^2}{4G},
\end{eqnarray}
where $m=m_0+\s$ is the constitute mass of quarks and $\s=-2G\lan\jb\j\ran$ and $\p^0=-2G\lan\jb i\g_5\t_3\j\ran$ represent the sigma and neutral pion condensates. The quark fields can then be integrated out analytically, which is most conveniently done in Euclidean path integral and the resultant thermodynamic potential reads
\begin{eqnarray}
\O=\frac{(m-m_0)^2+(\p^0)^2}{4G}-\frac{1}{V_4}\Tr\ln S^{-1},
\end{eqnarray}
where $V_4$ is the Euclidean spacetime volume, $\Tr$ acts on all physical spaces, and  the quark propagator is
\begin{eqnarray}
S(x,x')=-(i\slashed{D}-m-i\p^0\g_5\t_3)^{-1}\d^{(4)}(x-x').
\end{eqnarray}
It is diagonal in flavor space which allows us to calculate it analytically by adopting the Schwinger proper-time formalism~\cite{Schwinger:1951nm}. The result is
\begin{widetext}
\begin{eqnarray}
{\cal S}_{f}(x,x')
&=&{-i\over(4\pi)^2}\int_0^\infty {ds\over s^2}\;e^{-iq_{f}\int_{x'}^xA\cdot dx}
\big[-{1\over2}\gamma\big(q_{f}F\coth(q_{f}Fs)-q_{f}F\big)(x-x')+m-{\rm sgn}(q_{f})i\gamma^5\pi^0\big]\nonumber\\
&&\times\exp\Big\{-i\big[m^2+(\pi^0)^2\big]s+{i\over4}(x-x')q_{f}F\coth(q_{f}Fs)(x-x')+{i\over2}q_{f}\sigma Fs\Big\}{-(q_{f}s)^2I_2\over\text{Im}\cosh\big(iq_{f}s(I_1+2iI_2)^{1/2}\big)},\nonumber
\end{eqnarray}
\end{widetext}
where $f=u,d$, $I_1=\bB^2-\bE^2$ is the first Lorentz invariant, and we have suppressed some of the Lorentz scripts, e.g., $\s F$ should be understood as $\s^{\m\n}F_{\n\m}$ with $\s^{\m\n}=\frac{i}{2}[\g^\m,\g^\n]$. The true vacuum corresponds to the global minimum of $\O$. In the case that $I_1=0$ and $I_2>0$, we have numerically minimized $\O$ and obtained the corresponding $m$ and $\p^0$, see \fig{fig_m_pi0}. Once $I_2$ is turned on, finite $\p^0$ condensate arises and sigma condensate is suppressed --- a feature representing a chiral rotation from $\s$-direction to $\p^0$-direction; when $I_2^{1/4}\gtrsim 0.28 $ GeV, the CSB is driven by $\p^0$ condensate solely. In addition, the total condensate $m^*=\sqrt{m^2+(\p^0)^2}$ is weakened and finally killed by $I_2$ due to the presence of the electric field~\cite{Babansky:1997zh,Cohen:2007bt}.

To gain more insights, we write down the gap equations which is valid when $|I_2|<I_2^c$ (see below):
\begin{widetext}
\begin{eqnarray}
{m-m_0\over 2G}
&=&\frac{mN_c}{4\p^2}\sum_{f=u,d}\int_0^\infty ds e^{-\big[m^2+(\pi^0)^2\big]s}\;{q_{f}^2I_2{\rm Re}\cosh[q_f s\sqrt{I_1+2iI_2}]
\over{\rm Im}\cosh[q_f s\sqrt{I_1+2iI_2}]}-{N_c\over4\pi^{2}}{\pi^0\over{m^2+(\pi^0)^2}}
(q_u^2-q_d^2)I_2,\label{mgap}\\
{\pi^0\over 2G}
&=&\frac{\pi^0N_c}{4\p^2}\sum_{f=u,d}\int_0^\infty ds e^{-\big[m^2+(\pi^0)^2\big]s}\;{q_{f}^2I_2{\rm Re}\cosh[q_f s\sqrt{I_1+2iI_2}]
\over{\rm Im}\cosh[q_f s\sqrt{I_1+2iI_2}]}+{N_c\over4\pi^{2}}{m\over{m^2+(\pi^0)^2}}
(q_u^2-q_d^2)I_2.\label{pi0gap}
\end{eqnarray}
\end{widetext}
By eliminating $m$, we analytically obtain
\begin{eqnarray}
\label{njlpi}
\p_0=\frac{N_c}{4\p^2}\frac{2G}{m_0}(q^2_u-q^2_d)I_2.
\end{eqnarray}
By assuming the NJL version of the Gell-Mann-Oakes-Renner relation, $m^2_\p f_\p^2=m_0m^*(2G)^{-1}$~\cite{Klevansky:1992qe,GOR}, \eq{njlpi} immediately reduces to \eq{pi0cond} for $|I_2|<I_2^c$ which is independent of the model parameters $G$ and $\L$.

Two comments: (1) If $m_0=0$ or $m_\p=0$, the above formula is divergent. This corresponds to the chiral limit case where once $I_2$ is turned on the CSB is immediately rotated to the $\p^0$-direction. (2) When $|I_2|>I_2^c$, one of the gap equations is not applicable because $m$ is already zero. This situation is consistent with the ChPT result. In both NJL and ChPT frameworks, when $|I_2|>I_2^c$, the CSB is found to be totally driven by $\p^0$ condensation.

\section{Stability of the $\p^0$-condensed vacuum}
The neutral pion condensation found above requires a nonzero electric field. As we know, strong electric field can efficiently generate particle-antiparticle pairs (in the confined phase, mostly $\p^+$ and $\p^-$; in the deconfined phase, quarks and antiquarks) through the Schwinger mechanism and drive the vacuum unstable. So it is necessary to analyze that under which condition the $\p^0$-condensed vacuum is stable. For this purpose, we need to calculate the pair creation rate. This can be achieved within the ChPT framework by using the Schwinger proper-time method, see, e.g., Refs.~\cite{Kim:2003qp,Cohen:2007bt}. We will not repeat their calculations here, but just quote their results relevant to our discussion.

In case that $I_2\neq0$, we can always find a frame in which $\bE$ and $\bB$ are parallel. We will work in such a frame. The imaginary part of the leading order effective Lagrangian, which is responsible for the $\p^+\p^-$ production, reads
\begin{eqnarray}
\im \cl\!=\!\frac{e^2|EB|}{16\p^2}\!\!\!\sum_{n,l=1}^\infty\!\!\! \frac{(\!-\!1)^{n+1}}{n}\exp\!\!\ls\!-n\p\frac{m_\p^2\!+\!|eB|(2l\!-\!1)}{|eE|}\rs,\nonumber\\
\end{eqnarray}
where $n$ runs over all Schwinger poles and $l$ runs over all Landau levels.
The first term in the summation over $n$ defines the $\p^+\p^-$ production rate per unit volume which reads,
\begin{eqnarray}
w_{\p^+\p^-}= \frac{e^2|EB|}{16\p^2}\sum_{l=1}^\infty \exp\ls-\p\frac{m_\p^2+|eB|(2l-1)}{|eE|}\rs.
\end{eqnarray}
It is clear that the presence of the magnetic field effectively enhances the mass of charged pions,
\begin{eqnarray}
m^2_{\p^\pm}=m^2_\p+|eB|,
\end{eqnarray}
and therefore as long as
\begin{eqnarray}
\label{stability}
|eE|\ll m^2_\p+|eB|,
\end{eqnarray}
the rate of $\p^+\p^-$ production is strongly suppressed. In this case, the pair-producing instability takes place over a very long time $t\sim 1/(w_{\p^+\p^-} V)$ ($V$ the volume) and it is meaningful to consider the ``static" property of the vacuum. The ``equilibrium state" neutral pion condensate we have studied so far is sensible when the condition (\ref{stability}) is satisfied.

\section{Summary}
In this article, we propose a mean to tune the QCD vacuum by applying parallel electric and magnetic fields. The underlying mechanism is the electromagnetic triangle anomaly, through which the second Lorentz invariant $I_2=\bE\cdot\bB$ of the EM field can induce a neutral pion condensate in QCD vacuum and lead to a chiral rotation from the isosinglet scalar $\s$-direction toward the isotriplet pseudoscalar $\p^0$-direction. By adopting the chiral perturbation theory at leading order plus the Wess-Zumino-Witten term and the Nambu--Jona-Lasinio model at the mean-field approximation (leading order in $1/N_c$ expansion), we show that the sine of the rotation angle is universally given by \eq{pi0cond}.For weak electric field, our results may be testable in lattice QCD by adopting similar strategies as the ones used in lattice simulations at finite baryon chemical potential~\cite{DElia:2012zw}.

Our finding may have experimental implications in heavy-ion collisions. In ultra-peripheral heavy-ion collisions the hot quark-gluon matter may not form but quite large $I_2$ with opposite signs above and below the reaction plane may exist~\cite{Deng:2012pc} which can induce transient $\p^0$ condensation. The diphoton spectrum may be used to distinguish the $\p^0$-condensed vacuum from the normal vacuum, because in the $\p^0$-condensed vacuum the $\s$ and $\p^0$ excitations are mixed which will change the diphoton emission rate and spectrum. We will study these experimental signatures in near future.

\emph{Acknowledgments}---
We thank T. Brauner for very useful comments. GC and XGH are supported by Shanghai Natural Science Foundation (Grant No. 14ZR1403000), the Key Laboratory of Quark and Lepton Physics (MOE) of CCNU (Grant No. QLPL20122), the Young 1000 Talents Program of China, and Scientific Research Foundation of State Education Ministry for Returned Scholars. GC is
also supported by China Postdoctoral Science Foundation (Grant No. KLH1512072).


\begin{thebibliography}{20}
\bibitem{footnote}
For three flavor case, $M=\diag(m_0,m_0,m_s)$ and $Q/e=\diag(2/3,-1/3,-1/3)$. In this case, $Q/e$ is a linear combination of $\l_3$ and $\l_8$ and thus anti-commutes with other Gell-Mann matrices. In the limit $m_s\ra m_0$, the presence of a pure electric field or a pure magnetic field does not violate the positivity. But when $m_s\neq m_0$ or when both $\bE$ and $\bB$ are present, the positivity is spoiled.
\bibitem{GOR}
Note that in the presence of the EM field, this NJL version of the Gell-Mann-Oakes-Renner relation only holds approximately, because $m^*$ is EM-field dependent. However, from \fig{fig_m_pi0} we see that, for $|I_2|\lesssim I^c_2$, this approximation is quite good because in this case $m^*$ varies very little with $I_2$.
\end{thebibliography}
\end{document}